# Liquid crystal Skyrmions can swim

Paul J. Ackerman,[1,2] Timothy Boyle[1] and Ivan I. Smalyukh[1,2,3,4*]

[1] Department of Physics, University of Colorado, Boulder, CO 80309, USA
[2] Department of Electrical, Computer, and Energy Engineering, University of Colorado, Boulder, CO 80309, USA
[3] Soft Materials Research Center and Materials Science and Engineering Program, University of Colorado, Boulder, CO 80309, USA
[4] Renewable and Sustainable Energy Institute, National Renewable Energy Laboratory and University of Colorado, Boulder, CO 80309, USA
*Correspondence to: ivan.smalyukh@colorado.edu

**Abstract.** *Topologically nontrivial field configurations called "baby skyrmions" behave like particles and give origins to the field of skyrmionics that promises racetrack memory and other technological applications. Unraveling the non-equilibrium behavior of such topological solitons is a challenge. We realize baby skyrmions in a chiral liquid crystal and, using numerical modeling and polarized video microscopy, demonstrate electrically driven squirming motion. We reveal the intricate details of non-equilibrium topology-preserving textural changes driving this behavior. Direction of the skyrmion's motion is robustly controlled in a plane orthogonal to the applied field and can be reversed by varying frequency. Our findings may spur a new paradigm of soliton dynamics in soft matter, with a rich interplay between topology, chirality, and orientational viscoelasticity.*

Motion of topological defects enables the plasticity of metals and mediates condensed matter phase transitions, as well as is responsible for numerous irreversible processes in physical systems ranging from cosmology to elementary particles and condensed matter.[1] Topological defects enable the existence of thermodynamically stable condensed matter phases, such as the twist grain boundary and blue phases of liquid crystals (LCs).[1] However, apart from such exotic condensed phases that exhibit topological defects in the ground state, singular defects tend to annihilate and typically exist within transient short-lived processes. In soft active matter,[2] where weak external stimuli or local energy conversion fuel strong localized responses, unexpected recent findings include activity-driven generation and propulsion of the singular defects, spatially localized regions where order cannot be defined.[2-5] Active matter defects continuously progress through the cycles of generation and annihilation, displaying directional motion and other unexpected dynamics.[3-5] In chiral ferromagnetic solids, on the other hand, much of the recent interest is attracted by nonsingular solitonic two-dimensional (2D) topological defects,[6-12] which are often called "baby skyrmions" because they are low-dimensional topological counterparts of the higher-dimensional Skyrme solitons originally used to model elementary particles with different Baryon numbers.[13] These baby skyrmions can be stable both as individual particle-like excitations[7,11,12] and as building blocks of thermodynamically stable phases, such as the so-called "A-phase".[6,8] Furthermore, electric currents induce directional motion of such solitonic defects in solid films,[11] which is of interest for a new breed of racetrack memory devices and other skyrmionics applications.[12] However, the experimental study of field-induced dynamics of baby skyrmions presents a challenge because techniques capable of imaging spin textures and field configurations lack the needed simultaneous spatial and temporal resolution.[6-12]

In this work, we realize baby skyrmions in chiral nematic LCs, which were previously studied only as static field configurations,[14,15] and show that low voltages can drive motion of such topological defects in directions orthogonal to the applied electric field. The chirality-stabilized skyrmions "swim" like defects in active matter,[2-5] albeit never annihilating and being driven by modulation of applied voltage rather than local energy conversion. At the same time, similar to the skyrmions in ferromagnetic solids,[11,12] these skyrmions move without relying on fluid flows. Enabled mainly by rotational dynamics of the LC molecular alignment field, this low-Reynolds-number translational motion of baby skyrmions stems from electrically driven squirming of their localized twisted regions. Though these topology-protected twisted regions have no membrane or physical interface, periodic relaxation and tightening of the twisted region make the skyrmions expand, contract and morph as they move, in this aspect resembling squirming motion of biological cells[16] and biomimetic robots.[17] Applied voltage driving schemes control both the direction and speed of these topological defects uniformly throughout the sample. Using a combination of laser tweezers, video microscopy, three-dimensional nonlinear optical imaging and numerical modeling,[14,18] we uncover the underlying physical mechanisms and reveal connections to the intrinsic chirality and orientational viscoelasticity of the LC. The controlled motion of non-annihilating topological solitons may enable versatile forms of particle transport and dynamic memory devices,[12] and may potentially even spur a paradigm of out-of-equilibrium solitonic matter in LC systems.[2-5]

**Results**

**Experimental realization and numerical modeling of baby skyrmions.** We use a chiral nematic LC that has an intrinsic tendency for its local average rod-like molecular orientation **n**, dubbed "director", to twist around a helical axis orthogonal to **n**,[1] where $2\pi$ rotation of **n** defines a spatial periodicity called "helicoidal pitch" $p$. Our left-handed chiral LC is a nematic host MLC-6609 doped with a chiral additive ZLI-811 (both from Merck), with a negative dielectric anisotropy, such that **n(r)** tends to orient orthogonally to the applied electric field. The LC is confined in a glass cell of thickness $d \approx p$ with substrates treated to produce perpendicular boundary conditions for **n**. This confinement is incompatible with the helicoidal configuration of the chiral nematic LC and prompts unwinding of the director to a uniform vertically aligned state, as well as the formation of localized **n(r)**-configurations that embed energetically favorable twist while conforming to the uniform vertical far-field background $\mathbf{n}_0$ and the surface boundary conditions.[14,15] Figure 1 illustrates the geometry of our experiments, in which the skyrmion is oriented orthogonally to confining substrates and along the electric field applied to the transparent electrodes at the inner surfaces of the confining glass plates. When alternating current (AC) voltage with frequency $f_c$=1-10 kHz is applied to the electrodes, because of the LC's negative dielectric anisotropy, the director around the skyrmion rotates from its initial vertical orientation (Fig. 1a) to an in-plane orientation (Fig. 1b), whereas the skyrmion morphs its structure while staying oriented orthogonally to confining substrates. The direction of in-plane orientation of $\mathbf{n}_0$ in the cell midplane far away from skyrmions is either selected spontaneously as a result of a process (typically 5-10s long) involving coarsening of domains that follows voltage application, or it can be pre-determined by inducing a small pretilt with respect to the perpendicualr surface boundary conditions, as detailed in the methods section. When applied voltage is additionally modulated with lower modulation frequency $f_m < f_c$ (Fig. 1c), the skyrmon translates in a lateral direction orthogonal to $\mathbf{n}_0$ that depends on $f_m$ (along one of the black arrows depicted in Fig. 1c). The design of our voltage-driving scheme with high $f_c$ allows us to preclude complex hydrodynamic effects[1] that could be associated with the response of ionic impurities to low-frequency electric fields, which assures purely dielectric response of the LC to fields and is key for elucidating the physical mechanism behind the studied skyrmion dynamics.

Figure 2a shows an axially symmetric solitonic field configuration, as viewed using polarizing optical microscopy (POM) between crossed polarizers, with the corresponding results of detailed numerical modeling and nonlinear optical imaging[18] presented in Figs. 3 and 4, respectively.[19] A spontaneous formation of these solitons occurs during cooling of the sample through the isotropic-nematic phase transition, albeit they can be also controllably generated using laser tweezers.[14,15] With increasing the applied voltage above a well defined threshold ≈2V, the director far away from the baby skyrmion transitions from the initial vertical to tilted and then in-plane orientation. This director reorientation in response to the applied field is similar to that in LC displays.[20] This rotation morphs **n(r)** of the localized solitonic structure (Fig. 2b,c) to meet tilted or in-plane orientation of the far-field director $\mathbf{n}_0$ while minimizing the total free energy. However, the topologically protected solitonic structure withstands this dramatic realignment of the $\mathbf{n}_0$-background by ≈90°. The initial axially symmetric solitonic configuration (Fig. 2a) morphs to become asymmetric with respect to a vertical plane parallel to the new in-plane orientation of electrically rotated $\mathbf{n}_0$ (Fig. 2c,f). To gain insight into this field-driven transformation of baby skyrmions, we use numerical modeling of their equilibrium structures by minimizing the total free energy $W$ of the chiral nematic LC via the relaxation method:[14,21]

$$W = \int \left\{ \frac{K_{11}}{2} (\nabla \cdot \boldsymbol{n})^2 + \frac{K_{22}}{2} [\boldsymbol{n} \cdot (\nabla \times \boldsymbol{n}) + q_0]^2 + \frac{K_{33}}{2} [\boldsymbol{n} \times (\nabla \times \boldsymbol{n})]^2 - \frac{\varepsilon_0 \Delta \varepsilon}{2} (\mathbf{E} \cdot \boldsymbol{n})^2 \right\} dV \quad (1)$$

where $q_0=2\pi/p$ is the equilibrium chiral wavenumber, $K_{11}$, $K_{22}$ and $K_{33}$ are Frank elastic constants that pertain to splay, twist and bend distortions of **n(r)**, respectively, $\Delta\varepsilon$ is dielectric anisotropy (Supplementary Table 1)[19] and integration is over the sample's volume. Since the boundary conditions at the confining substrates used in experiments are strong and always satisfied,[14] W contains only the bulk terms of free energy and the surface energy does not enter this minimization problem (we note that the baby skyrmions can be also realized in cells with weak perpendicular boundary conditions[15], but this type of cell confinement is outside the scope of the present study). The energy-minimizing configurations of baby skyrmions (Fig. 2d-f), which emerge from a competition of LC elasticity and the dielectric coupling of **n** with electric field **E**, are consistent with the POM micrographs (Fig. 2a-c).

Figure 3 shows the nonpolar (Fig. 3a-d) and vectorized (Fig. 3f-i) representations of numerically simulated equilibrium director structures of the studied solitons before and after applying the electric field. These computer-simulated director structures are consistent with the experimental reconstruction of the director field. To demonstrate this, Fig. 4 compares experimental and computer-simulated three-photon excitation fluorescence polarizing microscopy (3PEF-PM) images corresponding to the director structures depicted in Fig. 3 for the case of circular polarization of the 3PEF-PM excitation light.[18] The good agreement between the experiment and modeling supports our interpretation of the experimental findings. Furthermore, a similar good agreement between experimental (Fig. 2g,h) and computer-simulated (Supplementary Fig. 1) structures is found also for the baby skyrmion oriented with its axis orthogonal to $\mathbf{n}_0$ and the translationally invariant 2D plane containing $\mathbf{n}_0$.

**Topology of baby skyrmions in chiral LCs.** At no fields, the baby skyrmion features a π-twist of $\mathbf{n}(\mathbf{r})$[22,23] from its center to periphery in all radial directions, smoothly embedded in the $\mathbf{n}_0$ background. Characterization of topology and energetic stability of these static localized field configurations was a subject of many recent studies,[14-15,21,24-27] so we discuss here only the topological aspects essential to understanding the electrically-driven transformation and dynamics of $\mathbf{n}(\mathbf{r})$-structures that we study in this work. We decorate $\mathbf{n}(\mathbf{r})$ and $\mathbf{n}_0$ with vectors to analyze the skyrmion topology.[22,23] Mapping the vectorized $\mathbf{n}(\mathbf{r})$ from the sample's midplane to the corresponding order parameter space, the two-sphere $S^2$ describing all possible orientations of a unit vector $\mathbf{n}(\mathbf{r})$, covers the $S^2$ once (Fig. 2a,d), consistent with the topologically protected nature of an elementary baby skyrmions.[10] This topology of the solitons remains unchanged at all used voltages (Fig. 2a-f) and is characterized by the same skyrmion number, an element of the second homotopy group $\pi_2(S^2)=Z$, despite the dramatic field-driven transformation of the localized solitonic structures. Because of the actual nonpolar nature of the LC director, $\mathbf{n}(\mathbf{r})\equiv-\mathbf{n}(\mathbf{r})$ and, similar to the case of point defects,[23] each elementary soliton that we study is characterized by a skyrmion number ±1,[14,15,21] with the sign determined only upon defining the vecorization direction. The most widely known skyrmions have axisymmetric translationally invariant (along their axes) 2D field configurations, like the one shown in Fig. 2d. However, the intense recent studies of skyrmionic structures in chiral LCs and ferromagnets revealed also non-axisymmetric skyrmions[28] and skyrmions with the effective cross-section shrinking or expanding along their length in response to boundary conditions, fields and other factors[14,15,24-32], and even skyrmions terminating, branching or nucleating at point defects.[31,32] Topological solitons with larger skyrmion numbers and also with nonzero Hopf indices can be realized in confined chiral nematic LCs as well,[15,21] but here we focus on the structures characterized by zero Hopf index and elementary skyrmion numbers ±1. The baby skyrmion in our case is matched to the perpendicular surface boundary conditions of the LC cell by the two hyperbolic point defects of opposite hedgehog charges occurring near the confining surfaces, forming the structure of the so-called "toron".[14] One can think about this structure as a baby skyrmion terminating near the cell substrates, with the termination points corresponding to the point defects (Fig. 3a,c,e,g).[21,31] This observation is natural as the $\pi_2(S^2)=Z$ skyrmions are topologically distinct from the uniform field defined along $\mathbf{n}_0$ by the surface boundary conditions and, in order to match these conditions, have to be terminated at their two ends by point defects of opposite hedgehog charge $\pi_2(S^2)=Z$,[21,31] which in our case is ±1.

The applied field morphs the director structure but retains its topology (Fig. 3b,d,f,h), including the skyrmion number of the topological soliton and the hedgehog charges of point defects near the confining surfaces. Despite different structural details, baby skyrmions with the two different orientations (having the 2D planes parallel and orthogonal to $\mathbf{n}_0$) are characterized by the same skyrmion number ±1, which we confirm by mapping the vectorized director field from both the experimental and computer-simulated 2D cross-sections of these structures orthogonal to the axes of skyrmions to the $S^2$ order parameter space (Fig. 2d-f). Smooth changes of $\mathbf{n}_0$ in response to fields do not alter topology of such elementary skyrmionic structures but continuously rotate $\mathbf{n}_0$ relative to the 2D plane of the baby skyrmion and the LC cell's midplane from orthogonal (Figs. 1a and 2a,d) to parallel (Figs. 1 and 2c,f). Such two mutual orientations of $\mathbf{n}_0$ and the skyrmion's 2D plane, orthogonal (Fig. 2a,d) and parallel (Fig. 2g-i), are also observed in our experiments at no fields (supplementary Fig. 1),[19] as well as have been previously observed in solid-state ferromagnets.[30] We track positions of spatial regions with vectorized $\mathbf{n}(\mathbf{r})$ corresponding to the north and south poles of $S^2$ (dubbed "preimages" of these $S^2$-points) versus voltage on the basis of POM video microscopy. This yields a linear dependence of the separation between the preimages of the north and south poles of $S^2$ versus voltage, with a good agreement between experimental and simulated dependencies and POM images (Fig. 5a,b).

**Brownian motion of baby skyrmions.** Baby skyrmions undergo Brownian motion, which resembles that of colloidal particles (Fig. 5c,d), consistent with their particle-like nature. Lateral diffusion of the skyrmions is driven by unbalanced orientational thermal fluctuations of $\mathbf{n}(\mathbf{r})$ within their localized structure[1] and is resisted by the LC's orientational viscous drag associated with the rotational viscosity $\gamma$.[1] Interestingly, such diffusion is voltage-dependent (Fig. 5d) along the **y**-axis diffusion but not along the **x**-axis (Fig. 5c), which is due to anisotropic morphing of these localized structures in applied field. By fitting the experimental data with Gaussians, we determine half-widths $\Delta$ of histograms of displacements and calculate both the lateral diffusivity of baby skyrmions $D=\Delta^2/\tau$ and the effective viscous drag coefficients according to the Einstein relation $\zeta=k_\mathrm{B}T/D$, where $k_\mathrm{B}=1.38\times10^{-23}$ J/K is Boltzmann's constant, $\tau=67$ms is the elapsed time corresponding to the frame rate of video microscopy, and T is the absolute temperature. At no fields, the viscous drag $\zeta \approx 2.3 \times 10^{-7}$ kg/s is direction-independent, but it becomes anisotropic and voltage-dependent at $U>2$V, with its values within $\zeta_x$=2.2-2.4$\times 10^{-7}$ kg/s for **x**-directions and $\zeta_y$=0.7-2.2$\times 10^{-7}$ kg/s for **y**-directions (Supplementary Table 2).[19]

**Electrically driven directional motion of the skyrmions.** An interesting dynamic emerges in response to changes of applied fields. The skyrmion drifts away from its original position along the **x**-direction when the voltage is switched "on" and drifts in the opposite –**x**-direction (more quickly, but to a smaller overall displacement) when the voltage is turned off (Fig. 6a). A net result of turning voltage on and off is that the baby skyrmion translates laterally because its fore–aft displacement in response to the electric switching is asymmetric with respect to the initial position. The effective Reynolds number associated with this highly overdamped motion can be estimated as Re$\approx 10^{-8} \ll 1$.[1,33,34] The in-plane force inducing the skyrmion motion (Fig. 6b), which emerges from the combination of LC elasticity and dielectric response, is balanced by the viscous drag force $F_d=-\zeta d\mathbf{r}/dt$, similar to that of Brownian motion emerging from rotational thermal fluctuations of **n**(**r**,t) within the soliton (Fig. 5c,d). Using the drag coefficients $\zeta$ determined on the basis of Brownian motion (Supplementary Table 2),[19] we find that the maximum in-plane forces $F \approx -F_d$ generated by electric pulses (balanced by the viscous drag) are in the pN-range (Fig. 6b). A sequence of electric pulses propels a skyrmion in a well-defined direction and can be exploited for navigating it throughout the LC sample. As an example, skyrmion displacements in response to a few cycles of turning voltage on and off are shown in Fig. 6c. By modulating the applied voltage we induce the net skyrmion motion in a direction along a vector connecting skyrmion's south- and north-pole preimages (along positive **x**) for the relatively large modulation periods $T_m$ (Fig. 6c, bottom inset) and in the opposite –**x** direction for smaller $T_m$. This behavior is consistent with the difference between the time-dependent |Δr| values corresponding to $T_m$ measured in response to turning voltage on and off (Fig. 6a). By varying $f_m=1/T_m$ within 0.3-100 Hz, the motion direction can be reversed and the skyrmion velocity can be varied from −0.2 to 2 μm/s (Fig. 6d). The skyrmion moves along **x** for $f_m$<5 Hz and along –**x** for $f_m$>5 Hz (Figs. 1c and 6d).

One might suspect that the observed voltage-controlled directional motion of skyrmions could involve the actual LC fluid flows and could be related to hydrodynamic effects, electroconvection, backflows or related phenomena. However, by using high $f_c$=1-10 kHz carrier frequencies, we eliminate the possible role that could be played by ionic currents and associated hydrodynamics. Moreover, by using plasmonic tracer nanoparticles directly observed using dark field videomicroscopy, we confirm that the observed motion cannot be associated with the LC fluid flows, such as backflows[1,33] (Fig. 6c). Using holographic laser tweezers, we position the plasmonic nanoparticles at different locations across the sample depth and in proximity or far away from the skyrmions. For the voltage-driving scheme and applied voltages used in our study, tracer nanoparticles exhibit no well defined directional motion (Fig. 6c), though the backflow effects at 4V cause barely noticeable shifting of nanoparticles back and forth as voltage is modulated, with the maximum observed effect shown in Fig. 6c while being compared to the corresponding lateral translations of skyrmions. Even though these weak backflows are detectable (Fig. 6c) and result in self-compensating back-forth translations of nanoparticles somewhat stronger than the Brownian motion of the same nanoparticles, they cannot account for the origin of the field-driven dynamics of skyrmions (the direction of these weak backflows changes across the cell thicknes, so that they roughly self-compensate and do not alter dynamics skyrmion motion). Although our tracer nanoparticles interact with the solitons, similar to the case of our previous studies,[35] these interactions do not reveal fluid flows in the skyrmion vicinity, additionally confirming that electroconvection and hydrodynamics effects cannot be responsible for the skyrmion motion. These findings pose a challenge of unraveling the physical origins of the observed soliton dynamics.

To gain insights to the physics behind the directional skyrmion motion, we again resort to numerical modeling. As we modulate the voltage applied to studied samples, a viscous torque associated with rotational viscosity $\gamma$ opposes the fast rotation of **n** in LCs in response to the competing electric and elastic torques.[1] The resulting director dynamics is governed by a torque balance equation,[1] $\gamma \partial n_i/\partial t = -\delta W/\delta n_i$, from which we obtain the temporal evolution $n_i(t)$ towards the equilibrium. As a result of the periodic modulation of voltage $U$, **n**(**r**,t) evolves in such a way that the computer simulated baby skyrmion translates, similar to what we observe experimentally (Figs. 6 and 7), albeit modeling at $f_m$<2 Hz is challenged by the large separation of north- and south-pole preimages during voltage modulation, which make calculations computationally very costly. Our modeling reproduces the fine features of the jerky motion of a baby skyrmion (top inset of Fig. 6c), the modulation-frequency dependence of its velocity (inset of Fig. 6d), and even the temporal evolution of experimental POM textures (Fig. 7a,b). In the sample midplane, the non-reciprocal squirming evolution of **n**(**r**,t) of the soliton within a period $T_m$ of modulating voltage at frequencies $f_m$=1 Hz (Fig. 7c) and $f_m$=20 Hz (Fig. 7d) is consistent with the opposite motion directions (Fig. 6d). To show this, we illustrate the evolution of preimages of the north-pole, south-pole, and equatorial regions of $S^2$, as well as the 2D handedness[21,36] $\mathcal{H}=-\mathbf{n}\cdot(\nabla\times\mathbf{n})$ describing twist of $\mathbf{n}(\mathbf{r})$ within the 2D plane of a baby skyrmion in a quantitative way (Fig. 7c,d). Interestingly, the high-$\mathcal{H}$ twisted region (colored in magenta) is morphing within $T_m$ in a way that resembles the traveling-wave surface squirming that biological cells[16] and biomimetic robots[17] use to swim in low Reynolds number fluid (Fig. 7c,d). The twisted region is squeezed and relaxed as voltage is applied and turned off within $T_m$, but the combination of chirality and coupling of $\mathbf{n}(\mathbf{r})$ to the modulated field breaks reciprocity of $\mathbf{n}(\mathbf{r},t)$, resulting in a directional motion of the soliton. The twisted high-$\mathcal{H}$ region of the baby skyrmion can be thought of as a 2D quasiparticle. The surface waves in this high-$\mathcal{H}$ twisted quasiparticle resemble the squirming traveling waves,[16,17] though they have very different origin as compared to biological and biomimetic systems. Importantly, this squirming motion involves only rotational dynamics of $\mathbf{n}(\mathbf{r})$ and not the actual LC fluid flow, although strong "backflows" can be induced at high voltages[33] and may provide a means of further enriching the out-of-equilibrium behavior of this solitonic system. Similar to active colloidal particles,[2] which locally convert chemical energy to self-propel, skyrmions locally convert electric energy due to the modulated applied voltage to produce motion by invoking the rotational dynamics of $\mathbf{n}(\mathbf{r},t)$. However, one should also understand the limitations of this analogy as compared to active particles and biological cells because, being just the localized field configurations, our skyrmions have no cell walls or physical interfaces.

Figure 8 shows a series of frames extracted from a POM video demonstrating that not only individual baby skyrmions exhibit directional motion, but also that multiple skyrmions within the same sample can exhibit such behavior. Interestingly, the dynamics of these different skyrmions is characterized by comparable velocity and the same motion direction (Fig. 8). Furthermore, as the number density of baby skyrmions increases, they start interacting with each other, which further enriches the non-equilibrium behavior of this solitonic systems and leads to the forms of self-assembly not accessible to the equilibrium structures. Detailed explorations of the non-equilibrium processes of such solitonic systems at large number densities of solitons will be discussed elsewhere.

**Discussion.**

Several recent studied have highlighted the topological, energetic and other analogies between LC and ferromagnetic solid-state solitons.[15,21,24,26,37] Indeed, within the one-constant approximation $K=K_{11}=K_{22}=K_{33}$ and for $A=K/2$, $D=Kq_0$ and $\mathbf{n}(\mathbf{r}) \rightarrow \mathbf{m}(\mathbf{r})$, the first three elastic terms of Eq.(1) reduce to become analogous to the simplest description of chiral ferromagnetic solids like MnSi[15,21,24,26,37]:

$$W = \int \{A(\nabla \mathbf{m})^2 + D[\mathbf{m} \cdot (\nabla \times \mathbf{m})]\} \, dV \quad (2)$$

This may provide insights into the physical underpinnings of certain types of dynamics of baby skyrmions in the spin textures and magnetization fields $\mathbf{m}(\mathbf{r})$ of the solid ferromagnetic films, though the coupling of $\mathbf{m}(\mathbf{r})$ to external electric and magnetic fields can be different from the dielectric coupling exploited in this work. By applying the $K=K_{11}=K_{22}=K_{33}$ one-constant approximation in our numerical modeling, we indeed find that the field-driven directional motion of skyrmions persists, though accounting for the LC's actual elastic constants is important for achieving the quantitative agreement between experimental and computer-simulated voltage-dependent structures and motion velocities (Figs. 4-7). Since the solitonic structures can also be realized in the fluid analogs of solid ferromagnetic films that combine properties of optical anisotropy with a facile magnetic response,[37,43] this may enable racetrack memory applications based on both magnetic and electric driving of skyrmion motion and both magnetic and optical data storage. Indeed, our preliminary studies show that periodically modulated magnetic fields applied to chiral ferromagnetic LC colloids with the localized skyrmionic similar to the ones studied here[37,43] also induce directional motion that stems from the structural squirming mechanism, similar to the one reported here. Furthermore, since the textures in LCs[1] and fluid ferromagnets[37] interact with colloidal microparticles because of the LC elasticity and boundary conditions for the director on surfaces of particles, their dynamics can be potentially exploited for the delivery of cargo on micrometer scales.[38]

From a fundamental standpoint of view, our study is the first demonstration of non-annihilating dynamics of $\pi_2(S^2)=Z$ topological solitons, very different from previously explored dynamics of both solitons and singular defects in LCs.[39-42] Experimental and theoretical studies of such non-equilibrium behavior may be also extended to skyrmions with large skyrmion numbers[15] and to the $\pi_3(S^2)=Z$ topological solitons with linked preimages and nonzero Hopf indices.[21,37] The directional motion of solitons can be also realized in materials with positive dielectric anisotropy[19] and also in response to magnetic fields, light, and other external stimuli. For example, to drive skyrmion motion with magnetic fields one can explore both the diamagnetic response of conventional LCs[1] (with a quadratic free energy term describing the coupling between **n** and magnetic field **B**), or also the ferromagnetic response of chiral LC ferromagnets,[37] in which both the $\pi_2(S^2)=Z$ and $\pi_3(S^2)=Z$ topological solitons have been recently realized.[37,43]

To conclude, we have uncovered unexpected directional motion of topological solitons in response to modulated electric fields. This finding bridges the recent key developments in studies of out-of-equilibrium phenomena[2-5,17,44] and topological solitons,[6-15,21,37] and may lead to versatile topology-protected dynamic systems with potential applications in microfluidics, racetrack memory devices, etc. Directional motion of solitons is enabled by non-reciprocal squirming of their particle-like textures and involves only rotational dynamics of the LC director, without actual fluid flow, establishing the particle-like out-of-equilibrium solitonic structures as a new kind of stimuli-responsive soft particles. Our observations and numerical analysis may provide insights into the means of inducing directional motion of skyrmionic textures in ferromagnetic solids, with potential impact on the emergent field of skyrmionics.[12]

**Methods.**

**Details of Sample Preparation and Experimental Techniques.** The LC material for which we report majority of experimental and computational studies is a mixture of a nematic host MLC-6609 and a chiral additive ZLI-811 (both purchased from Merck) with an equilibrium helicoidal pitch $p \approx 10$ μm. This LC mixture was designed to have negative dielectric anisotropy and respond to external electric field in such a way that the director tends to orient orthogonally to the applied field direction. The material parameters of this LC system, including dielectric and elastic constants, rotational viscosity and refractive indices, are provided in the Supplementary Table 1. In addition, test experiments were also done for several other material systems (Supplementary Table 1), as briefly discussed below. In some samples, small number densities (about one particle per 10,000 µm$^3$) of plasmonic gold nanorods with dimensions about 22 × 50 nm were added to the LC for tracing possible fluid flows on the basis of observing the nanoparticle motion using dark field videomicroscopy; the preparation of gold nanorods and their dilute dispersions in LCs is described in details elsewhere.[46]

LC cells were assembled using glass substrates with transparent indium tin oxide (ITO) electrodes treated with polyimide SE1211 (purchased from Nissan) for imposing surface boundary conditions. SE1211 was spin-coated at 2700 rpm for 30 s and then baked for 5 min at 90 °C followed by 1 h baking at 180 °C to provide strong vertical surface boundary conditions for the LC director. For some of the studied samples, by weakly rubbing SE1211-coated glass plates with initially perpendicular boundary conditions for **n**, we have defined a small (1-3°) pretilt away from these perpendicular conditions (see Ref.[45] for more details), so that the direction of tilting of **n** during electric switching could be well defined, which was then pre-determining the direction of skyrmion motion as discussed above. The LC cell gap was set to be 10 µm using glass-fiber segments dispersed in ultraviolet-curable glue. Small drops of glue with the spacers were squeezed between the glass substrates with inward-facing ITO electrodes and alignment layers and the glue was then cured using ultraviolet exposure to obtain LC cells, fixing the desired cell gap. The LC material was infiltrated to the cells by means of capillary forces and then the cell edges were sealed with 5-min fast-setting epoxy. Electric wires were soldered to the ITO electrodes and used to apply voltages. Different voltage driving schemes and waveforms were produced using a homemade LabView-based software and the data acquisition board (NIDAQ-6363, National Instruments) and applied to the LC cells using the wires soldered to the ITO. An amplitude modulated 1 kHz square wave was used to avoid hydrodynamic instabilities and other types of complex behavior associated with ions at low-frequency applied fields. The voltage driving-scheme was based on the $f_c$=1 kHz square-wave carrier frequency modulated by a lower frequency $f_m$ square-wave. The relevant parameters of the electric voltage driving are illustrated in the bottom inset of Fig. 6c.

Non-contact generation and manipulation of solitonic configurations was achieved using holographic laser tweezers and a tightly focused 1064 nm laser at powers of less than 50 mW, as described in detail elsewhere.[14,15] For this, we utilized an Ytterbium-doped fiber laser (YLR-10-1064, IPG Photonics, operating at 1064 nm) and a phase-only spatial light modulator (P512-1064, Boulder Nonlinear Systems) integrated into a holographic laser tweezers setup capable of producing arbitrary 3D patterns of laser light intensity within the sample.[14,15] The laser tweezers were also integrated with the 3D nonlinear optical imaging setup described below, enabling fully optical generation, control, and nondestructive imaging of the solitons. The physical mechanism behind the laser generation of solitons is the optical Fréedericksz transition, the realignment of the LC director away from the far-field background $\mathbf{n}_0$ caused by its coupling to the optical-frequency electric field of the laser beam, which is described by a corresponding term of free energy.[14,15] This coupling, enriched by holographically generated patterns of the trapping laser beam's intensity, phase singularities, and translational motion of individual traps,[14,15] prompts complex director distortions that relax to the global or local elastic free energy minima, some of which are the baby skyrmions we study in this work.

Polarizing optical microscopy observations in the transmission mode were done by using a multi-modal imaging setup built around the IX-81 Olympus inverted microscope (also includes the nonlinear optical imaging technique described below) and charge coupled device cameras. POM textures of undulations and colloidal particles were recorded with a camera Flea FMVU-13S2C-CS (purchased from Point Grey Research, Inc.) or a Spot 14.2 Color Mosaic (purchased from Diagnostic Instruments, Inc.) or a fast camera HotShot 512SC (purchased from NAC Image Technology, Inc.). The same high numerical aperture (NA) Olympus objective (100×, NA=1.4) was used for the polarizing optical microscopy, laser manipulation and nonlinear optical imaging described below. Optical imaging additionally utilized 10×, 20×, and 50× dry objectives with numerical aperture NA=0.3-0.9. The sequences of video microscopy frames were analyzed using ImageJ software and its plugins (freeware from NIH), which allowed for tracking lateral positions of the north-pole and south-pole preimages (which both appear dark in the POM images) within the baby skyrmions with 7-10 nm resolution and the temporal resolution corresponding to the camera's frame rates, which was 1000 frames per second for the fast camera HotShot 512SC.

Nonlinear optical imaging was performed using three-photon excitation fluorescence polarizing microscopy (3PEF-PM) setup built around a IX-81 Olympus inverted optical microscope.[21] The polarized self-fluorescence from the LC molecules was detected within the 400-450 nm spectral range and excited through a process of three-photon absorption using a Ti-Sapphire oscillator (Chameleon Ultra II, Coherent) operating at 870 nm with 140 fs pulses at a repetition rate of 80 MHz. The 3PEF-PM signal was collected through an oil-immersion 100× objective with numerical aperture of 1.4 and detected by a photomultiplier tube (H5784-20, Hamamatsu). We scanned the excitation beam through the sample volume with the help of galvano-mirrors (in lateral directions) and a stepper motor (across the sample thickness) and recorded the 3PEF-PM signal as a function of coordinates, which was then used to construct 3D images by means of the ParaView software (freeware obtained from KitwarePublic).[21] The linear polarization of the excitation beam was controlled using a polarizer and a rotatable half-wave retardation plate. The detection channel utilized no polarizers. The 3PEF-PM intensity scaled as $\propto \cos^6\psi$, where $\psi$ is the angle between $\mathbf{n}(\mathbf{r})$ and the excitation beam's linear polarization[18,21] (assumed to remain unchanged despite the beam focusing through dielectric interfaces and the weakly birefringent LC medium, with the sample design minimizing these changes) and was used to reconstruct the $\mathbf{n}(\mathbf{r})$ patterns.[18,21] The reconstruction of the 3D solitonic $\mathbf{n}(\mathbf{r})$-structures took advantage of the self-fluorescence patterns obtained at different polarizations of excitation light, as described elsewhere.[18,21] In order to eliminate the ambiguity between the two possible opposite $\mathbf{n}(\mathbf{r})$ tilts in the analysis of 3D images, additional cross-sectional 3PEF-PM images were obtained at orientations of the LC cell's normal tilted by ±2° with respect to the microscope axis for linear polarizations of excitation laser light parallel or perpendicular to the plane of the corresponding vertical cross-sectional image. The $\mathbf{n}(\mathbf{r})$ tilt ambiguity was then eliminated based on the $\propto \cos^6\psi$ scaling of the 3PEF-PM signal and the ensuing spatial changes of intensity prompted by the ±2° tilts. To further narrow the angular sector of $\mathbf{n}$-orientations corresponding to preimages of points on $S^2$ with target azimuthal angles $\varphi$, we each time obtained three 3D images with azimuthal orientation of the linear polarization of excitation beam at φ and φ±3°. These 3D images were smoothed using Matlab-based software and then used in a differential analysis to improve orientational resolution of imaging the director field to better than ±3°.

**Details of Numerical Modeling.** In computer simulations, three-dimensional grids of $112 \times 112 \times 32$ equally spaced points were used with periodic boundary conditions along lateral edges of the computational volume and perpendicular boundary conditions at the top and bottom planes.[14,21,47] The utilized values of the elastic, dielectric, optical and viscous constants are summarized in the Supplementary Table 1. To obtain the equilibrium director structures at different applied fields, starting from random initial conditions, **n(r)**-configurations were relaxed using a centered finite difference scheme to calculate the functional derivatives $\delta W/\delta n_i$ of the free energy given by Eq. (1), where $n_i$ is the component of **n** along the $i^{th}$ axis ($i$ = **x**, **y**, **z**). The maximum stable time step $\Delta t$ was used to change the components of **n** by an elementary displacement calculated as described in detail elsewhere.[14,21] The new **n** was normalized after each time step. This process was repeated until the grid averaged elementary displacement is less than $10^{-14}$. Director dynamics was modeled using the torque balance equation, $\gamma \partial n_i/\partial t = -\delta W/\delta n_i$, from which we obtain the temporal evolution $n_i(t)$ towards the equilibrium at a given applied voltage $U$. The computer-simulated director structures of baby skyrmions were then used in modeling of polarizing optical micrographs by means of the Jones-matrix approach for the experimental material and cell parameters, such as optical anisotropy, equilibrium cholesteric pitch, and cell thickness (Supplementary Table 1).

The **n(r)**-configurations in the vertical cross-sections of the solitonic structures were represented in the form of arrays of azimuthal and polar angles. To utilize the Jones-matrix method for modeling of POM textures, we split the cell into a stack of 32 thin sub-layers while assuming that the orientation of **n(r)** is constant across the thickness of one of these sub-layers. The corresponding coordinate-dependent Jones matrices had an optical axis defined by orientation of **n(r)** and the phase retardation defined by the optical anisotropy of the LC and polar angle of the director. The resulting polarizing optical micrographs were obtained as a result of successive multiplication of Jones matrices corresponding to a polarizer, a series of thin LC slabs each equivalent to a phase retardation plate with spatially varying optical axis and retardation, and an analyzer. A two-dimensional polarizing optical image (see examples in Fig. 7b) was then obtained by performing such a Jones-matrix calculation for each pixel and then composing a two-dimensional texture with coordinate-dependent intensity analogous to the experimental images (see examples in Fig. 7a). To properly account for the achromatic nature of our experimental POM observations, we calculated these textures separately for three different wavelengths spanning the entire visible spectrum (475 nm, 510 nm, and 650 nm) and then superimposed them according to the experimental light source intensities at the corresponding wavelengths. This yielded achromatic polarizing optical micrographs analogous to the corresponding experimental images (Fig. 7a,b). Computer-simulated 3PEF-PM cross-sections were constructed for the director structures obtained as descried above by first finding the coordinate-dependent angles $\psi$ between **n(r)** and the linear polarization of the probing laser light and then plotting the normalized signal intensity as $I_{\text{3PEF-PM}} = \cos^6\psi$ (note that this modeling disregards finite resolution effects, as well as various issues associated with scattering and defocusing of light in the birefringent medium). Despite the large number of simplifications and assumptions, all experimental and computer-simulated images of **n(r)**-structures in chiral nematic samples closely match each other and strongly support our interpretation of experimental findings (Figs. 4, 5a,b and 7a,b).

**Data availability** The material parameters are provided in Supplementary Information. Matlab and Fortran codes and both experimental and computational datasets generated and analyzed during the current study are available from the corresponding author on request.

**Supplementary Information** is available in the online version of the paper.

**Acknowledgements.** This research was supported by the National Science Foundation Grant DMR-1410735. We thank Noel Clark, Ghadah Sheetah, Hayley Sohn, Mykola Tasinkevych, Sumesh Thampi and Julia Yeomans for discussions.

**Author contributions.** P.J.A., T.J.B and I.I.S. contributed to all aspects of this work and wrote the manuscript. I.I.S. conceived and designed the project.

**Author information**. Reprints and permissions information is available at www.nature.com/reprints. The authors declare no competing financial interests. Readers are welcome to comment on the online version of this article at www.nature.com/nature. Correspondence and requests for materials should be addressed to I.I.S. (ivan.smalyukh@colorado.edu).

# Figures:

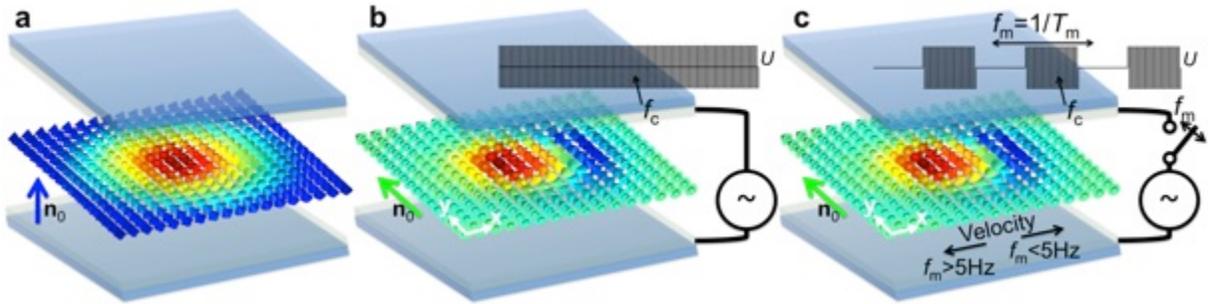

**Fig. 1 | Schematic illustration of experimental geometry. (a)** A skyrmion oriented with its axis orthogonal to confining substrates at no applied fields. **(b)** Electric field with frequency $f_c$ =1-10kHz applied across the sample thickness rotates the far-field director $\mathbf{n}_0$ away from the skyrmion to be in-plane and morphs the skyrmionic structure. **(c)** When this electric field with frequency $f_c$ =1-10kHz is additionally modulated with the modulation frequency $f_m$, the skyrmion moves in a direction determined by $f_m$.

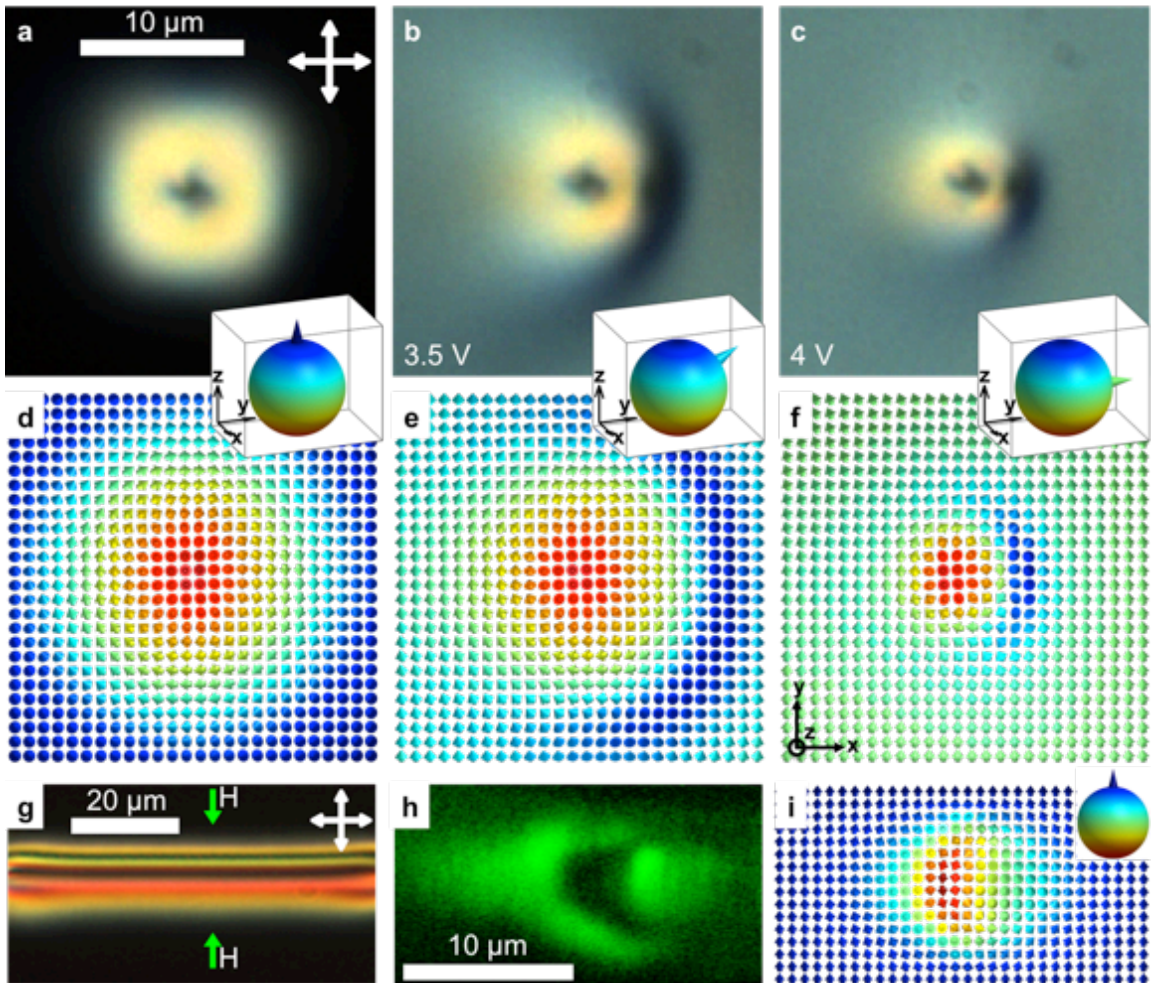

**Fig. 2 | Topology and electric switching of baby skyrmions in a confined chiral nematic LC. (a-c)** POM micrographs of a skyrmion at (a) no applied fields and (b,c) at voltages indicated on images. Crossed polarizers are shown using white double arrows in

(A). $f_c$=1kHz. (**d-f**) Computer-simulated vectorized **n(r)**-configurations corresponding to (a-c), where **n(r)** is shown using arrows colored according to their orientations and corresponding points on the $S^2$-sphere (insets), with the orientation of $\mathbf{n}_0$ on $S^2$ shown using cones. (**g**) POM micrograph of a baby skyrmion in a glass cell with perpendicular boundary conditions; the polarized nonlinear optical image of the skyrmion's 2D plane containing $\mathbf{n}_0$ is shown in (**h**) in the vertical cross-section along the H-H line of (g), with the corresponding computer-simulated **n(r)** depicted in (**i**); computer-simulated analogs of images (g,h) are shown in the Supplementary Fig. 1.[19]

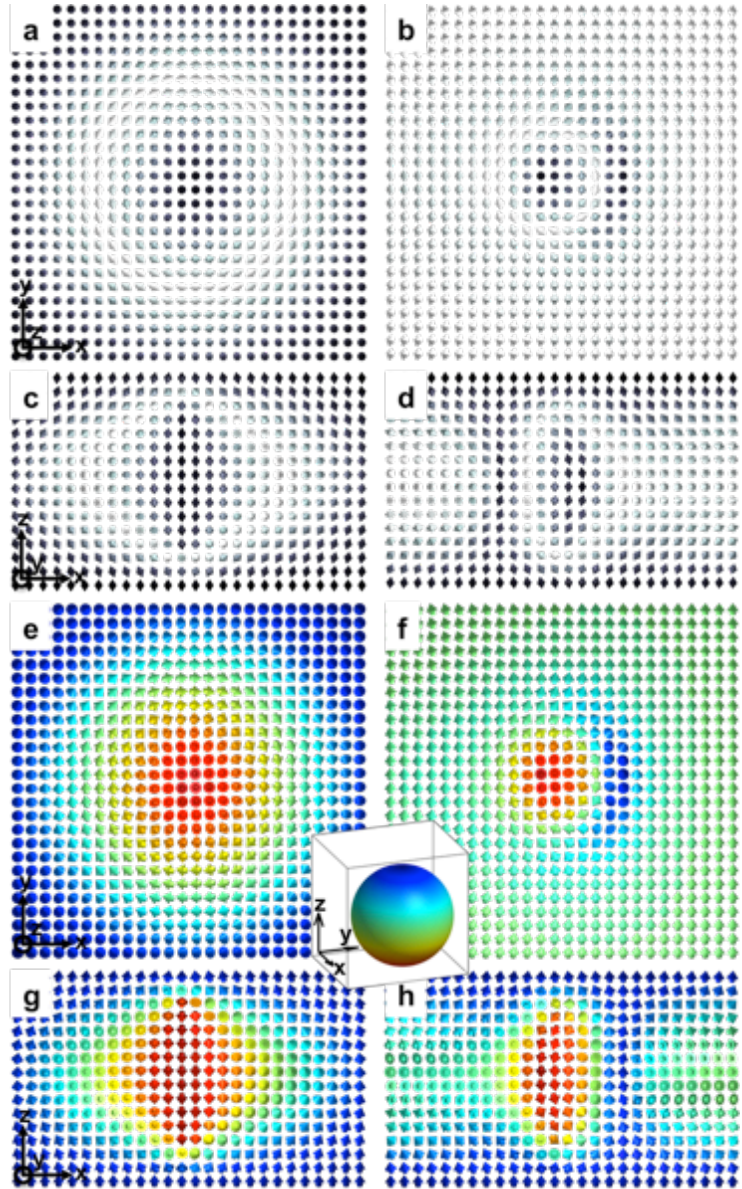

**Fig. 3 | Three-dimensional structure of baby skyrmions in a confined LC cell geometry.** (**a,b**) Director structure of the skyrmion in the cross-sectional plane orthogonal to $\mathbf{n}_0$ for (a) $U$=0 V and (b) $U$=4 V presented using double cones locally aligned with their axes along **n(r)**. (**c,d**) The corresponding director structure of the

skyrmion in the cross-sectional plane parallel to $\mathbf{n}_0$ and passing through the skyrmion's center for (c) $U$=0 V and (d) $U$=4 V. (**e-h**) The corresponding director structure of the skyrmion in the (e,f) in-plane and (g,h) vertical cross-sections visualized using arrows colored according to the corresponding points on $S^2$ (inset) and to the coloring scheme used in Fig. 2.

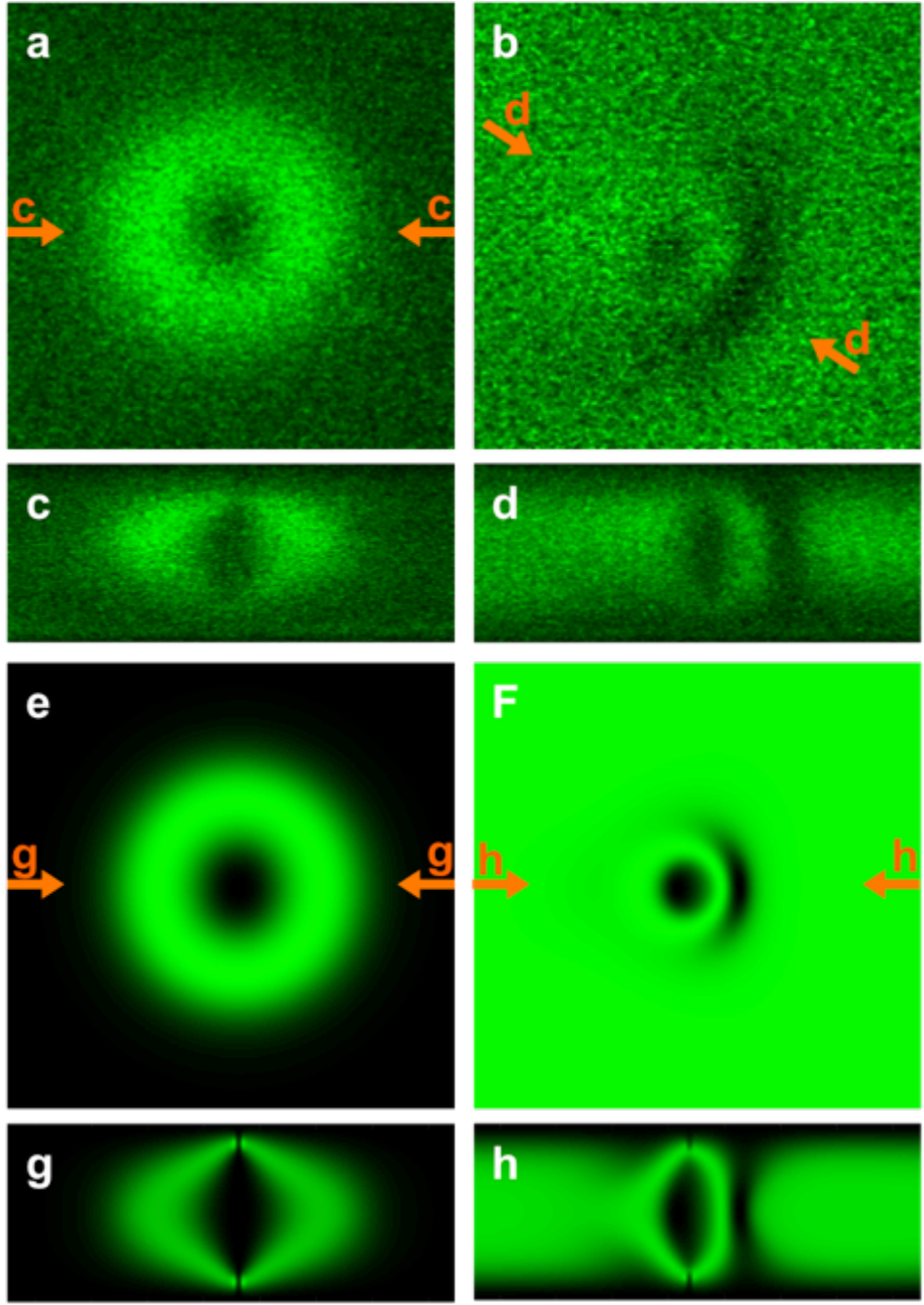

**Fig. 4 | 3PEF-PM images of baby skyrmions and their comparison with numerical modeling counterparts.** (**a**,**b**) Experimental 3PEF-PM images of the director structure of the skyrmion in the cross-sectional plane orthogonal to $\mathbf{n}_0$ for (a) $U$=0 V and (b) $U$=4 V. $f_c$=1kHz. (**c**,**d**) The corresponding experimental 3PEF-PM images of the director structure

of the-skyrmion in the vertical cross-sectional plane parallel to $\mathbf{n}_0$ and passing through the skyrmion's center for (c) $U=0$ V and (d) $U=4$ V. (**e-h**) The corresponding computer simulated 3PEF-PM images of the skyrmion in the (e,f) plane perpendicular to $\mathbf{n}_0$ and (g,h) vertical cross-sections parallel to $\mathbf{n}_0$. The 3PEF-PM excitation light used in imaging was circularly polarized.

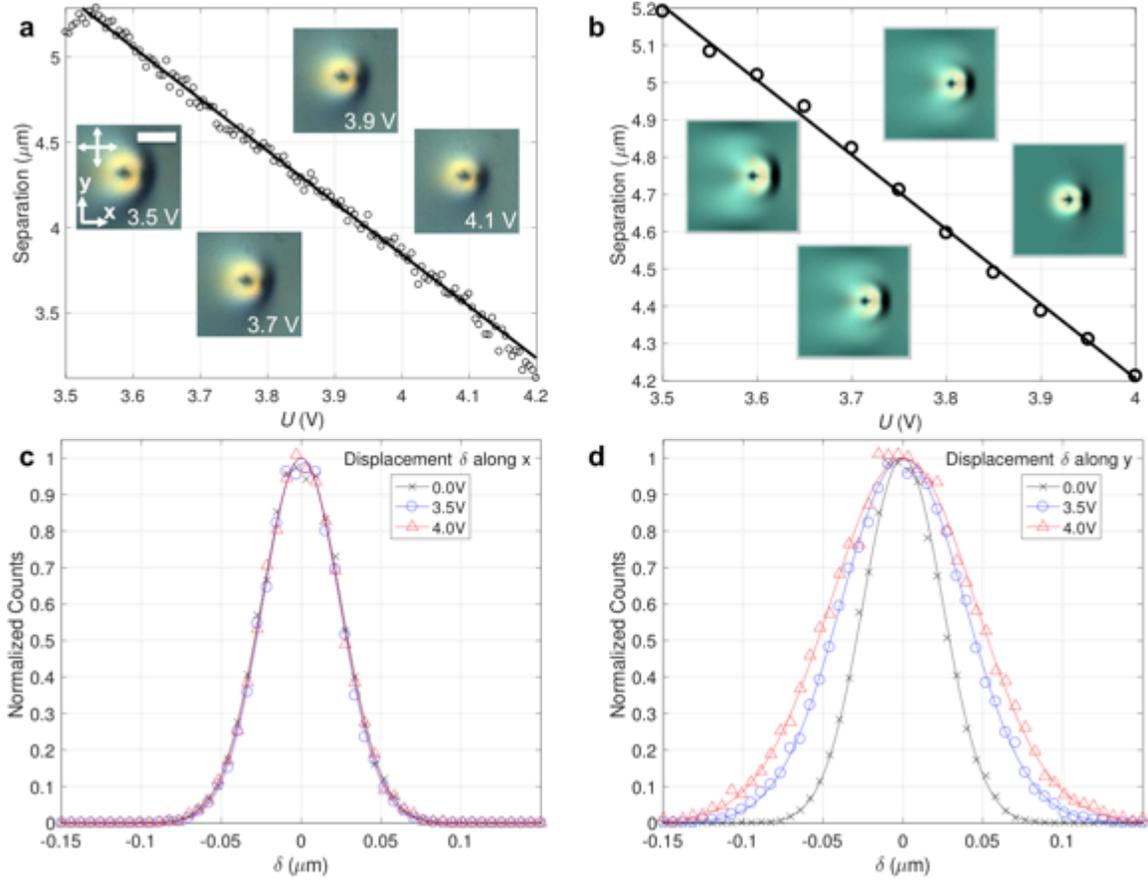

**Fig. 5 | Electric control and Brownian motion of baby skyrmions.** (**a,b**) Lateral separation between north- (blue arrows in Fig. 2d-f) and south-pole (red arrows) preimages characterized (a) using POM videos and (b) computer simulations. Insets show experimental and computer-simulated POM micrographs at voltages and for crossed polarizers (white double arrows) marked in (a). The scale bar in insets is 10 μm. $f_c$=1kHz. (**c,d**) Histograms of displacement $\delta$ of the south-pole preimage characterizing Brownian motion of a baby skyrmions (c) along the **x**-axis and (d) along the **y**-axis at different $U$; the solid lines are Gaussian fits of the experimental data points (symbols).

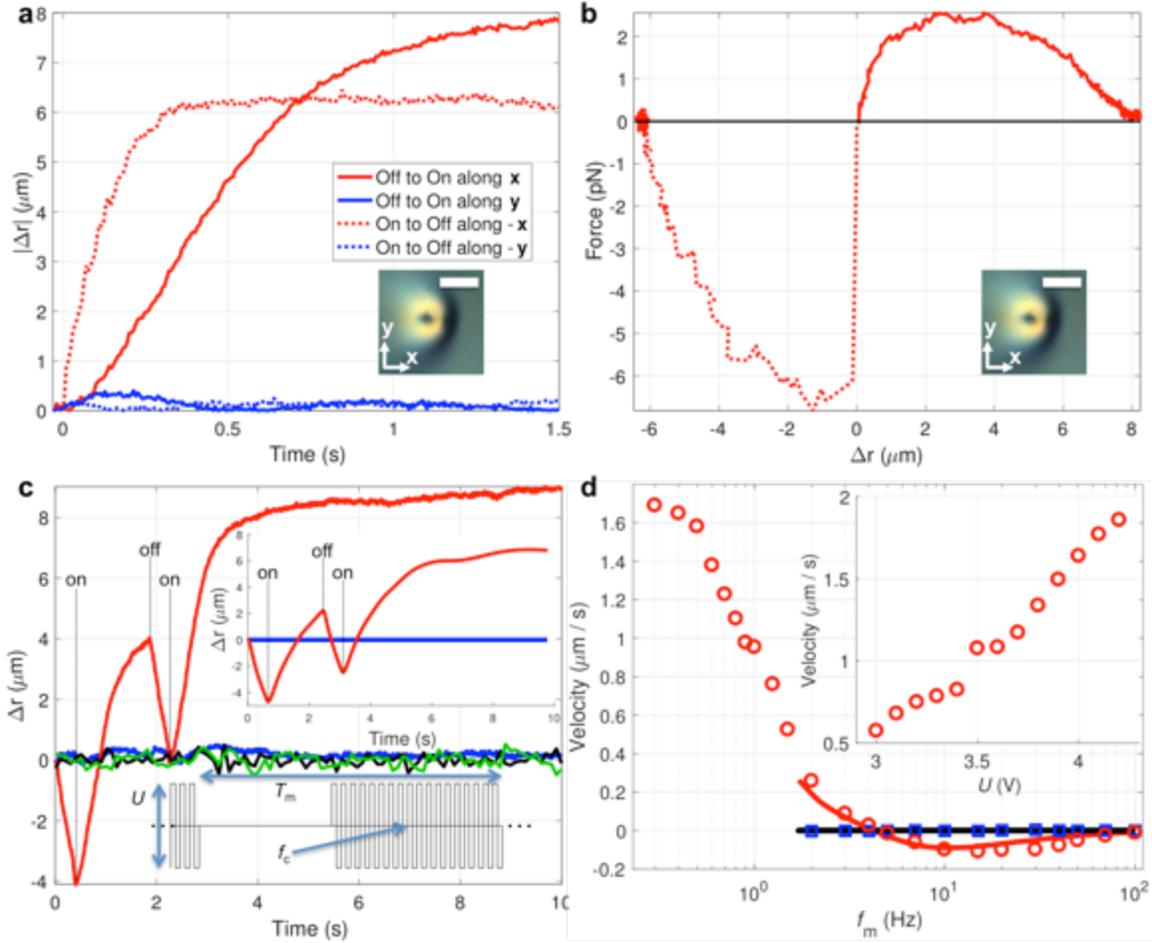

**Fig. 6 | Directional motion of baby skyrmions.** (**a**) Relative shift of spatial position of the skyrmion's south-pole preimage |Δr| versus time upon turning voltage $U$=4 V on (solid lines, motion along positive **x,y**) and off (dotted lines, motion along negative **x,y**) for the **x**- and **y**-directions defined in the inset and in Fig. 1f. (**b**) Corresponding forces along ±**x** (inset) acting on the baby skyrmion versus Δr. POM micrographs in the insets of (a,b) define the coordinate system; the scale bar is 10 μm. (**c**) Video microscopy characterization of the directional motion of the skyrmion in response to switching of the applied voltage on and off when starting with the voltage on; the corresponding computer simulated results are shown in the top inset. The bottom inset illustrates the square waveform voltage driving with carrier frequency $f_c$=1 kHz and modulation period $T_m$. The directional motion of the skyrmion along the **x**-axis is compared to the motion of a tracer nanoparticle along the same **x**-axis at no field (black solid line) and at an applied voltage of 4V (green solid line) with the same used voltage driving scheme and other conditions. (**d**) Dependence of the direction and amplitude of the baby skyrmion's velocity on the modulation frequency $f_m$=1/$T_m$=0.3-100 Hz characterized using video microscopy for a duty cycle of 75%; the experimental data are shown using symbols and the corresponding computer-simulated dependencies within $f_m$=1/$T_m$=2-100 Hz are shown using solid lines. Inset shows the skyrmion's velocity versus voltage.

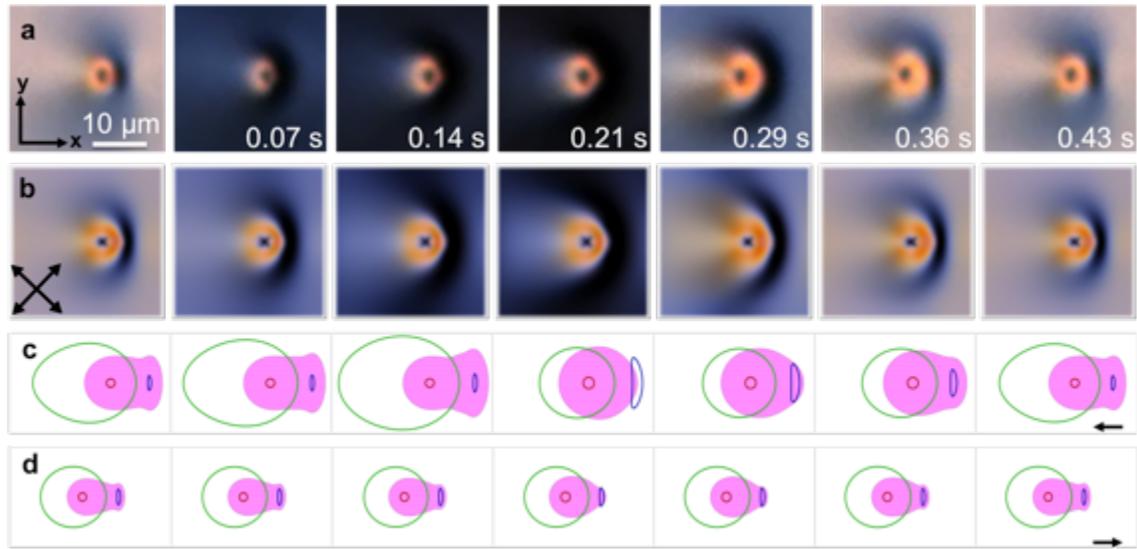

**Fig. 7 | Squirming of skyrmions probed using video microscopy and computer simulations.** (**a**) Experimental and (**b**) computer-simulated POM micrographs of a baby skyrmion when moving along a vector connecting the south- and north-pole preimages (along positive **x**), with the corresponding visualizations of squirming waves in the twisted quasi-particle illustrated in (**c**). Magenta filled surfaces depict the twisted regions of $\mathcal{H}/q_0 \geq 0.1$ of the chiral nematic LC. Crossed polarizers (black double arrows) in (b) correspond to both computer-simulated and experimental micrographs. A skyrmion is depicted for frames equally spaced in time during one cycle at $f_c$=1k Hz and $f_m$=2 Hz. After turning voltage off, the north-pole preimage expands and the south-pole preimage moves away from the north-pole preimage. Once the voltage is reinstated, the north-pole preimage localizes at a larger distance away from the south-pole preimage, and then the north- and south-pole preimages come back together within the laterally compact skyrmion that overall shifts along **x**, exhibiting directional motion. (**d**) Visualization of dynamics of a skyrmion quasiparticle similar to that shown in (c) but for $f_m$=20 Hz yielding an opposite direction of motion, where magenta filled surfaces depict the twisted regions with $\mathcal{H}/q_0 \geq 0.05$. In (c,d), the red and blue contours depict the spatial regions of the south- and north-pole preimages and the green contours show the cumulative preimage of the equatorial points of $S^2$. Black arrows in (c,d) show the skyrmion's motion directions at corresponding $f_m$.

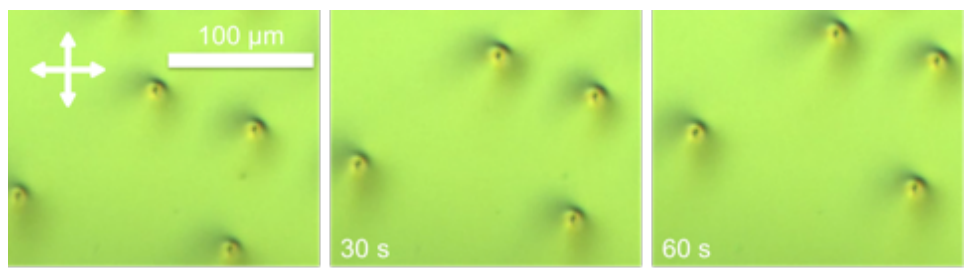

**Fig. 8 | Collective unidirectional motion of baby skyrmions shown using frames from a polarized microscopy video.** The elapsed time is marked in the bottom-left corners of the images. The orientation of crossed polarizers is shown using white double arrows.

# Supplementary Information

**Liquid crystal Skyrmions can swim**

Paul J. Ackerman,[1,2] Timothy Boyle[1] and Ivan I. Smalyukh[1,2,3,4*]

[1]*Department of Physics, University of Colorado, Boulder, CO 80309, USA*
[2]*Department of Electrical, Computer, and Energy Engineering, University of Colorado, Boulder, CO 80309, USA*
[3]*Soft Materials Research Center and Materials Science and Engineering Program, University of Colorado, Boulder, CO 80309, USA*
[4]*Renewable and Sustainable Energy Institute, National Renewable Energy Laboratory and University of Colorado, Boulder, CO 80309, USA*
*\*Correspondence to: ivan.smalyukh@colorado.edu*

**Diversity of material systems exhibiting skyrmion dynamics**

The topological soliton dynamics that we report in this article is not restricted to the particular nematic host material system of MLC-6609, for which we present most of the data, but can be achieved with a broad variety of LC materials with both positive and negative dielectric anisotropy. For example, similar experimental and numerical results are also obtained (to be reported elsewhere) using a chiral nematic LC based on a nematic mixture of ZLI-2806 (see material parameters in the Supplementary Table 1) doped with either a left-handed chiral additive ZLI-811, the same as that used in our current study, or a right-handed additive CB-15 (all materials purchased from Merck). Furthermore, translational motion of baby skyrmions shown in Fig. 2g-i and Supplementary Fig. 1 can be driven electrically even when obtained in materials with positive dielectric anisotropy, such as pentylcyanobiphenyl (5CB, purchased from Frinton Laboratories, Inc.), which we will describe elsewhere. In a material with positive dielectric anisotropy, such as 5CB, the field that is applied perpendicular to substrates squeezes the localized skyrmion structure (such as the one shown in the supplementary Fig. 1c) asymmetrically. When voltage is modulated (supplementary Fig. 1c), this too results in motion because of the field-driven asymmetric squirming response of the localized structure.

**Supplementary Tables:**

**Supplementary Table 1 |** Material parameters of the chiral nematic LC mixture MLC-6609 used in our studies, including dielectric and elastic constants, rotational viscosity and refractive indices. For, comparison, we also present material parameters of two other nematic hosts for which we also observed dynamics of skyrmionic structures, which will be reported elsewhere. The physical parameters presented in the table were measured at 20 ºC. The dielectric constant values are reported for the 1 kHz applied field and the refractive indices are provided for a monochromatic light at 589.3 nm wavelength.

| Material/parameters | MLC-6609 | ZLI-2806 | 5CB |
|---|---|---|---|
| $\varepsilon_\perp$ | 7.1 | 8.1 | 5.2 |
| $\varepsilon_\parallel$ | 3.4 | 3.3 | 19 |
| $\Delta\varepsilon$ | -3.7 | -4.8 | 13.8 |
| $K_{11}$ (pN) | 17.2 | 14.9 | 6.4 |
| $K_{22}$ (pN) | 7.51 | 7.9 | 3 |
| $K_{33}$ (pN) | 17.9 | 15.4 | 10 |
| $\gamma$ (mPas) | 162 | 240 | 77 |
| ne | 1.5514 | 1.518 | 1.726 |
| no | 1.4737 | 1.474 | 1.533 |
| $\Delta$n | 0.0777 | 0.044 | 0.193 |

**Supplementary Table 2 |** Lateral diffusivity and the effective viscous drag coefficients of baby skyrmions.

| | Along **x** | | Along **y** | |
|---|---|---|---|---|
| $U$ (V) | $D$ (μm²/s) | $\zeta$ (kg/s) | $D$ (μm²/s) | $\zeta$ (kg/s) |
| 0.0 | 0.0187 | $2.174\times10^{-7}$ | 0.0185 | $2.195\times10^{-7}$ |
| 3.5 | 0.0172 | $2.356\times10^{-7}$ | 0.0447 | $9.081\times10^{-8}$ |
| 4.0 | 0.0180 | $2.259\times10^{-7}$ | 0.0616 | $6.590\times10^{-8}$ |

**Supplementary Figures**

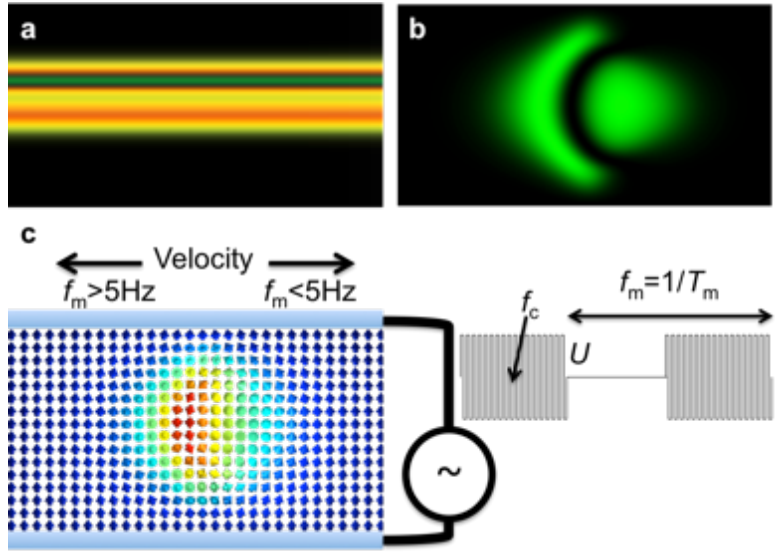

**Supplementary Figure 1 | Computer-simulated POM and 3PEF-PM images of baby skyrmions oriented orthogonally to $n_0$ at no fields, as well as geometry of samples.** (**a**) Computer-simulated analog of the POM image shown in Fig. 1g. (**b**) Computer-simulated analog of the 3PEF-PM image shown in Fig. 1h. These **n(r)**-structure and images correspond to one type of cholesteric fingers that we studied previously.[43,45] (**c**) A schematic showing **n(r)** and geometry of a sample in which translational motion of a baby skyrmion is induced in one of the two directions orthogonal to it, depending on $f_m$.

**Supplementary Videos:**

**Supplementary Video 1 | Reversible motion of a baby skyrmion.** The modulation frequency of applied modulated voltage $U$=4 V is $f_m$=1 Hz ($f_c$=1 kHz) in the first part of video (indicated in the bottom-left corners of the video frames), where the skyrmion moves from right to left, and then $f_m$=20 Hz in the second half of the video, when the skyrmion reverses its motion direction and moves from left to right. The video was sped up 15 times. The crossed polarizers in POM are along the video frame edges. The POM video was taken at a frame rate of 1 frame per second, so that the details of textural changes within $T_m$ cannot be resolved and only the ensuing directional motion can be seen. The video was taken using a camera Flea FMVU-13S2C-CS (from Point Grey Research, Inc.).

**Supplementary Video 2 | POM textural changes of the baby skyrmion during the directional swimming.** The modulation frequency of applied modulated voltage $U$=4 V is $f_m$=0.5 Hz ($f_c$=1 kHz). The video was sped up 2 times. The crossed polarizers in POM are along the video frame edges. The POM video was taken at a frame rate of 15 frames per second, so that the details of textural changes within $T_m$ can be clearly resolved, showing how they relate to the ensuing directional motion. The video was taken using a camera Flea FMVU-13S2C-CS (purchased from Point Grey Research, Inc.).